\documentclass[a4paper,11pt]{article}
\usepackage{jheppub} 
\usepackage{lineno}
\usepackage{amssymb}

\usepackage{graphicx} 
\usepackage{slashed}
\usepackage{tikz-cd}
\usepackage{tabularx,ragged2e}
\usepackage{amssymb,fge}
\usepackage{amsmath,amssymb}
\usepackage{slashed}
\usepackage{hyperref}
\usepackage{caption}
\usepackage{dsfont}
\usepackage{verbatim}
\usepackage{subfig}
\usepackage{mathtools,ytableau,amsfonts,tikz}
\usepackage{graphicx}
\usepackage{physics}

\title{Non-SUSY physics and the Atiyah-Singer index theorem}

\author[1]{Shunrui Li} 
\author[2,3]{and Yang Liu}

\affiliation[1]{Department of mathematics, Free University of Berlin, Berlin 14195, Germany}
\affiliation[2]{Department of Physics, Tsinghua University, Beijing 100084, China}
\affiliation[3]{School of Physics and Astronomy, University of Nottingham, University Park, Nottingham NG7 2RD, United Kingdom}

\emailAdd{lis02@zedat.fu-berlin.de}
\emailAdd{liu-yang\_1990@mail.tsinghua.edu.cn}

\abstract{
The Atiyah-Singer index theorem, a cornerstone of modern mathematics, has traditionally been derived from supersymmetric (SUSY) physics. This paper demonstrates a direct derivation from non-supersymmetric quantum statistics by establishing a fundamental correspondence: the grand partition functions of non-interacting bosonic and fermionic systems are precisely the Chern characters of certain vector bundles. Furthermore, we generalize this correspondence to infinite dimensions, where we construct a novel mathematical framework of spectral-sheaf pairs. Within this framework, we formulate a generalized index theorem, identifying the topological index with a regularized spectral product. This work not only circumvents the need for supersymmetry but also provides a deeper unifying perspective, revealing quantum statistics as a sufficient foundation for topological invariants.}

\begin{document}
\maketitle
\flushbottom

\section{Introduction}
\label{sec:intro}

The mathematical framework of quantum theories has been demonstrated to have a deep connection to global aspects of differential and algebraic geometry \cite{Witten1982a,SMc1967,Witten1982b,Witten1982c}. 

To begin, we review some fundamental aspects of supersymmetric theories. In a supersymmetric quantum field theory, one can introduce the quantity $\text{Tr}(-1)^F$ (where $F$ represents the fermion number) \cite{Witten1982a,SMc1967}. When appropriately regularized, this quantity counts the difference between the number of bosonic and fermionic states in the Hilbert space of the theory. The nature of supersymmetry, however, ensures that this index only depends on the zero-energy states, since all non-zero energy states appear in pairs of bosons and fermions. Moreover, $\text{Tr}(-1)^F$ remains invariant under continuous deformations of the Hamiltonian, making it a topological index of the entire quantum theory \cite{Witten1982a, SMc1967}. This topological property of $\text{Tr}(-1)^F$ was utilized in \cite{Witten1982a, SMc1967}, among other things, to connect the possibility of supersymmetry breaking in supersymmetric $\sigma$-models to the vanishing of certain topological invariants associated with the manifold on which the $\sigma$-model is defined.

In \cite{AG1983}, it was demonstrated that by considering quantum mechanical supersymmetric systems -- specifically, field theories in $0+1$ dimensions -- and using $\text{Tr}(-1)^F$ as a topological invariant, one can compute the index density for the Atiyah-Singer index theorem for classical complexes \cite{AS1968a, AS1968b, AS1968c, AS1971a, AS1971b, ABP1973, Gilkey1974}, along with their associated G-index generalizations. The findings in this paper are particularly striking, as they show that simple supersymmetric systems, such as the $N=1$ and $N=\frac{1}{2}$ supersymmetric $\sigma$-models, can provide significant topological insights into the manifold on which they are defined. It is important to note that the analysis in \cite{AG1983} assumes that the manifold to be compact and boundaryless.

Today, the index theorem derived from supersymmetry has become a fundamental research paradigm in mathematical physics. However, one may still pose the question: Can non-supersymmetric (non-SUSY) physics also yield the Atiyah-Singer index theorem? In \cite{Li:2025qry}, we demonstrated that a simple example -- the quantum harmonic oscillator -- can lead to the generalized Hirzebruch signature theorem, which is a special case of the Atiyah-Singer index theorem. In this paper, we extend the results presented in \cite{Li:2025qry}. We will show that a broad class of Atiyah-Singer index theorems can be derived from non-SUSY physics, specifically through the Bose-Einstein and Fermi-Dirac distributions.

The structure of the paper is organized as follows: in Section 2, we provide a brief review of the basic concepts that will be used throughout the paper. In Section 3, we demonstrate that the grand partition function can be interpreted as a special case of Chern character of sheaf. In Section 4, we “physically” derive the index theorem from non-interacting systems for vector bundles. In Section 5, we generalize the results in Section 4 to the infinite dimensional case and speculate some mathematical results for infinite dimensions. Finally, the conclusions and discussions are presented in Section 6. 

\section{Basics}

\subsection{The grand canonical ensemble and non-interacting system distribution}
In statistical mechanics, the grand canonical ensemble provides the fundamental framework for describing systems that can exchange both energy and particles with a reservoir. This is the natural ensemble for deriving the equilibrium distribution functions for quantum particles, which are inherently indistinguishable and whose behavior is governed by their quantum statistical nature: Bose-Einstein statistics for integer spin particles (bosons) and Fermi-Dirac statistics for half-integer spin particles (fermions).

\subsubsection{The grand canonical ensemble}
The grand canonical ensemble describes a system in equilibrium with a large reservoir at constant temperature $T$ and constant chemical potential $\mu$. The key thermodynamic potential is the grand potential, $\Omega$, defined as:
\begin{equation} \label{Omega}
\Omega = - k_B T \ln \Xi,
\end{equation}
where $k_B$ is the Boltzmann constant and $\Xi$ is the grand canonical partition function:
\begin{equation} \label{Xi1}
\Xi = \sum^{\infty}_{N=0} \sum_s e^{-\beta (E_s - \mu N)}.
\end{equation}
Here, $\beta=1/k_B T$, $N$ is the number of particles, and the sum over $s$ encompasses all microstates with energy $E_s$ for a fixed $N$. 

For a system of non-interacting particles, the total energy of a microstate is a sum of single-particle energies, $E_s = \sum_i n_i \epsilon_i$, where $\epsilon_i$ is the energy of the $i$-th single-particle state and $n_i$ is its occupation number. The total number of particles is $N= \sum_i n_i$. The constraint of summing over all $N$ and all microstates $s$ can be vastly simplified. Because the particles are non-interacting, the partition function factorizes. Crucially, the sum over all microstates for all particle numbers is equivalent to summing over all possible occupation numbers $n_i$ for every single-particle state $i$, with the appropriate statistical constraints. Therefore, the grand partition function can be rewritten as:
\begin{equation} \label{Xi2}
\Xi = \prod_i \Xi_i,
\end{equation}
where $\Xi_i$ is the partition function for the $i$-th single-particle state, summing over all possible ways to occupy that state.

The average occupation number of a state with energy $\epsilon_i$ is given by the general grand canonical formula:
\begin{equation} \label{ni1}
\langle n_i \rangle = - \frac{\partial}{\partial \alpha} \ln \Xi_i. 
\end{equation}
Thus, the problem reduces to the calculation $\Xi_i$ for a single state of energy $\epsilon_i$, with the specific rules for bosons and fermions determining the sum over $n_i$.

\subsubsection{Bose-Einstein distribution and Fermi-Dirac distribution}
For bosons, there is no restriction on the occupation number $n_i$ can be any integer from 0 to infinity. The single-state partition function is therefore a geometric series:
\begin{equation} \label{XiBEi1}
\Xi^{BE}_i = \sum^{\infty}_{n_i=0} \left[ e^{- (\alpha + \beta \epsilon_i)} \right]^{n_i} =  \sum^{\infty}_{n_i=0} \left( e^{-(\alpha + \beta \epsilon_i)} \right)^{n_i},
\end{equation}
where $\alpha=-\mu/(k_BT)$. This series converges only if $e^{-(\alpha + \beta \epsilon_i)} < 1$. Assuming this condition is met, then we have:
\begin{equation} \label{XiBEi2}
\Xi^{BE}_i =  \frac{1}{1- e^{-(\alpha + \beta \epsilon_i)}}.
\end{equation}
The average occupation number is:
\begin{equation} \label{niBE1}
\langle n_i \rangle^{BE} = -  \frac{\partial}{\partial \alpha} \ln \Xi^{BE}_i = -  \frac{\partial}{\partial \alpha} \ln \left(  \frac{1}{1- e^{-(\alpha + \beta \epsilon_i)}} \right). 
\end{equation}
Performing the differentiation yields the Bose-Einstein distribution:
\begin{equation} \label{niBE2}
\langle n_i \rangle^{BE} =   \frac{1}{e^{(\alpha + \beta \epsilon_i)}-1}. 
\end{equation}

For fermions, the Pauli exclusion principle dictates that each single-particle state can be occupied by at most one particle. Therefore, the occupation number $n_i$ can only be 0 or 1. The single-state partition function is:
\begin{equation} \label{XiFDi1}
\Xi^{FD}_i = \sum^{1}_{n_i=0}  e^{-(\alpha + \beta \epsilon_i) n_i}  = 1+  e^{-(\alpha + \beta \epsilon_i)}.
\end{equation}
The average occupation number is then:
\begin{equation} \label{niFD1}
\langle n_i \rangle^{FD} = - \frac{\partial}{\partial \alpha} \ln \Xi^{FD}_i = - \frac{\partial}{\partial \alpha} \ln \left(  1+ e^{-(\alpha + \beta \epsilon_i)} \right). 
\end{equation}
Performing the differentiation yields the Fermi-Dirac distribution:
\begin{equation} \label{niFD2}
\langle n_i \rangle^{FD} =   \frac{1}{e^{(\alpha + \beta \epsilon_i)}+1}. 
\end{equation}

If $e^{\alpha} \gg 1$, Bose-Einstein distribution and Fermi-Dirac distribution become Maxwell-Boltzmann distribution
\begin{equation} \label{niMB}
\langle n_i \rangle^{MB} =  e^{-(\alpha + \beta \epsilon_i)},
\end{equation}
which is the distribution for classical particle system. In Section 2.1.2, we have assumed that there is only one quantum state for every energy level $\epsilon_i$.

\subsection{Todd class and the Atiyah-Singer index theorem}

\subsubsection{Todd class}
In the intersection of algebraic geometry, differential topology, and index theory, characteristic classes provide powerful cohomological invariants that encode essential global properties of vector bundles and manifolds. Among these, the Todd class, denoted $Td(\mathcal{E})$, stands as a particularly pivotal construction. It is a rational cohomology class associated with a complex vector bundle $\mathcal{E}$ (or a complex manifold) that plays an indispensable role in the Hirzebruch-Riemann-Roch theorem and its far-reaching generalization, the Atiyah-Singer index theorem. While its definition is inherently topological, its consequences are deeply algebraic and geometric.

The Todd class is defined multiplicatively on complex vector bundles. For a complex vector bundle $\mathcal{F}$, its Todd class is given by the formal power series:
\begin{equation} \label{TdL}
Td(\mathcal{F}) = \prod_i \frac{x_i}{1- e^{-x_i}},
\end{equation}
where $x_i$ is a 2-form. 
While its definition is formal, the Todd class carries profound geometric meaning. It can be interpreted as a multiplicative sequence that measures the deviation of a complex manifold from being a product of complex projective spaces.

The Todd class is far more than a technical construction. It is a cornerstone of modern geometry. It serves as the crucial bridge linking the analytic world of elliptic operators and sheaf cohomology with the topological world of characteristic classes. It appeared in one of the most profound theorems of the twentieth century, Atiyah-Singer index theorem, highlighting its fundamental role in understanding the deep structure of complex manifolds and their bundles.

Its primary significance was established by Friedrich Hirzebruch. For a holomorphic vector bundle $E$ over a compact complex manifold $X$, the Hirzebruch-Riemann-Roch theorem states:
\begin{equation} \label{chi}
\chi (X,E) = \int_{X} ch(E) \cdot Td(X).
\end{equation}
Here, $\chi (X,E)= \sum^{\text{dim \ X}}_{i=0} (-1)^i \text{dim} H^i (X,E)$ is the holomorphic Euler characteristic, $ch(E)$ is the Chern character of the bundle $E$, and the integral denotes the evaluation of the top-dimensional cohomology class on the fundamental cycle of $X$. This theorem magnificently relates a fundamental algebraic invariant (the dimension of cohomology groups) to a topological invariant (the integral of a characteristic class). The Todd class $Td(X)$ is the precise correction factor that makes this remarkable equality hold \cite{ABP1973}.

\subsubsection{The Atiyah-Singer index theorem}

In mathematics, the Atiyah-Singer (AS) index theorem links analysis (solutions to differential equations) with topology (global properties of manifolds). Specifically, it relates the analytic index of an elliptic differential operator to its topological index. For example, for the Dirac operator on a manifold:
\begin{equation} \label{ASITDO}
\text{index}(D)= \text{dim} \ \text{ker} D -  \text{dim} \ \text{ker} D^{\dagger},
\end{equation}
where $D$ is the Dirac operator and $D^{\dagger}$ is its adjoint. This index equals a topological invariant, such as an integral involving curvature forms (like the $\hat{A}$ genus or Chern class) \cite{atiyah1963index}. 

The Atiyah–Singer theorem also has significant implications in theoretical physics. It explains the appearance of zero modes of fermions in certain backgrounds (like magnetic monopoles or instantons). It is crucial in understanding anomalies -- e.g., the chiral anomaly is connected to the index of the Dirac operator and provides a bridge between quantum theory and topology. Specially, the chiral anomaly is derived from the non-invariance of the fermionic path integral measure, which is deeply linked to the index of the Dirac operator on a curved or gauge field background. This makes the anomaly a topological effect, robust and independent of the specific details of the system \cite{nakahara2018geometry,nash1988topology,atiyah2006elliptic}.


\section{From single-particle to grand partition functions}
This section will explain in a simple and easy-to-understand way why the grant partition function of non-interacting boson and fermion systems can be realized as a Chern characters of natural sheaf-theoretic objects on the thermal spacetime \(M\times S^1_\beta\), where $M$ is $3$-dimensional Euclidean space and $S^1_\beta$ comes from compacting of time in thermal quantum field theory (In the thermal field theory, time is compacted to a circle with a circumference of $\beta$)\cite{Li:2025qry}. Let us explain the rationale for considering $M\times S^1_\beta$ from physical perspective and mathematical perspective. More details can be found in \cite{Li:2025qry}.

\paragraph{The physical motivation: incorporating finite temperature} 
The consideration of the product manifold $M\times S^1_\beta$ is fundamentally motivated by the need to incorporate finite-temperature effects into our topological framework. In thermal quantum field theory, the equilibrium properties of a quantum system at temperature $T$ are encoded by compactifying the imaginary time dimension with a periodicity equal to the inverse temperature, $\beta=1/(k_B T)$. This constructs the “thermal circle”, $S^1_{\beta}$. For a bosonic system, the fields satisfy periodic boundary conditions along this circle. The full spacetime is thus extended from the base manifold to the thermal spacetime $M\times S^1_\beta$. Physically, the path integral over this extended manifold computes the thermodynamic partition function, $Z=\text{tr} (e^{-\beta H})$. By formulating our theory on $M\times S^1_\beta$, we inherently place the quantum harmonic oscillator in a thermodynamic context, allowing its statistical mechanical properties, like the partition function and internal energy, to be treated within a geometric-topological formalism.

\paragraph{The mathematical framework: sheaves and characteristic classes} 
From a mathematical perspective, the transition to $M\times S^1_\beta$ provides the necessary geometric structure to reinterpret physical quantities as topological invariants. We introduce the concept of a “physical sheaf” $\mathcal{S}$, a Hermitian vector bundle over spacetime whose sections represent quantum states. On the original spacetime $M$, this sheaf describes the system at zero temperature. The extension to $M\times S^1_\beta$ is achieved via the pullback of this sheaf along the projection map $\pi: M\times S^1_\beta \rightarrow M$, defining a sheaf $\mathcal{S}_\beta := \pi^*\mathcal{S}_0$. This construction is natural from the viewpoint of algebraic geometry and facilitates the application of the Grothendieck-Riemann-Roch (GRR) theorem. The central result of this section is the invariance of the Chern character under this thermal compactification: 
\begin{equation}    
\sigma^*\mathrm{ch}(\mathcal{S}_\beta) = \mathrm{ch}(\mathcal{S}_0),
\end{equation}
where $\sigma: M \hookrightarrow M\times S^1_\beta $ is a “section map”. This invariance, a direct consequence of the functoriality of characteristic classes, rigorously establishes that the partition function $Z$, identified as the Chern character $\mathrm{ch}(\mathcal{S}_\beta)$, is a well-defined topological object even in the presence of a thermal circle.

Now we explain why the grant partition function of non-interacting boson and fermion systems can be realized as a Chern characters of natural sheaf-theoretic objects on the thermal spacetime \(M\times S^1_\beta\). We proceed from a single particle to the full system and provide short schematic diagrams to help readers unfamiliar with sheaf language.\\
\paragraph{1. Single particle} 
For a single particle, we fix a single-particle energy level labeled by $i$ with energy $\varepsilon_i$, and subtract the chemical potential to get $x_i=\varepsilon_i-\mu$. In thermal physics a single occupancy contributes a Boltzmann factor $e^{-\beta x_i}$ when it winds once around the thermal circle. 

We now convert these physical expressions into sheaf theory language. We fix a single-particle level \(i\) with energy \(\varepsilon_i\) and set \(x_i=\varepsilon_i-\mu\). Then we consider the \(i\)-th energy level as a rank-one object (a “line-like” sheaf) \(\mathcal S_i\) on \(M\times S^1_\beta\), the effect of one traversal of the thermal circle is encoded by the Chern root
\begin{equation}\label{Chern root}
    y_i \;=\; -\beta x_i,
\qquad\text{so that}\qquad \mathrm{CH}(\mathcal S_i)=e^{y_i}=e^{-\beta x_i}.
\end{equation}

We provide a schematic diagram of the above work:
\begin{equation}
    \begin{tikzcd}[row sep=small]
\text{level } i \arrow[r] & \mathcal S_i \arrow[r, "\mathrm{CH}"] & e^{y_i}=e^{-\beta x_i}.
\end{tikzcd}
\end{equation}

\paragraph{2. Bosonic system: symmetric algebra (local Fock Bosonic sheaf)} 
For a bosonic system the occupation number can be any nonnegative integer $n=0,1,2,\cdot \cdot \cdot$. The Fock space of one level is the direct sum of the symmetric tensor powers of the single-particle state. Therefore, we can give the definition of the $i$-th bosonic Fock sheaf $\mathcal S_i^B$ of the single-particle energy level $i$, which consists of components corresponding to all non-negative integers $n$: the $n$-th component is the $n$-particle subspace obtained by copying the single particle space $S_i$ $n$ times and symmetrizing the copies. These subspaces of different $n$ are pieced together by direct sum $\oplus$ to obtain the entire bosonic Fock sheaf $S_i^B$. The component $S_i^0=\mathcal O$ represent the case where there are no particles in energy level $i$ 
Therefore,
\begin{equation} \label{Def of SB}
    \mathcal S_i^B \coloneqq \operatorname{Sym}^\bullet(\mathcal S_i)=\bigoplus_{n\ge0}\mathcal S_i^{\otimes n}.
\end{equation}

Because of splitting principle of Chern characters, we get
\begin{equation}
    \mathrm{CH}(\mathcal S_i^{\otimes n})=\bigl(\mathrm{CH}(\mathcal S_i)\bigr)^n=e^{n y_i};
\end{equation}
then we sum over occupations to give the familiar single-level factor \footnote{The splitting principle allows reduction of bundle statements to the rank-one case by formally introducing Chern roots, representing each level by a rank-one object is natural and sufficient here. In addition, because the bosonic symmetric algebra is infinite, one often works with a complete symmetric algebra or treats $CH$ as a formal power series. In practice one chooses a regularization (zeta/heat-kernel) or requires convergence conditions.}
\begin{equation}
    \mathrm{CH}(\mathcal S_i^B)=\sum_{n\ge0} e^{n y_i}=\frac{1}{1-e^{y_i}}=\frac{1}{1-e^{-\beta x_i}}.
\end{equation}
We provide the schematic diagram for bosonic local Fock sheaf:
\begin{equation}
    \begin{tikzcd}[row sep=small]
\mathcal S_i \arrow[r, "\operatorname{Sym}^\bullet"] & \mathcal S_i^B \arrow[r, "\mathrm{CH}"] & \dfrac{1}{1-e^{y_i}}.
\end{tikzcd}
\end{equation}

\paragraph{3. Fermionic system: exterior algebra (local Fock sheaf)} 
A fermionic symstem allows \(n=0\) or \(1\). Therefore, we can represent the local fermionic Fock sheaf by the exterior algebra on \(\mathcal S_i\) (similar to the definition of the boson case):
\begin{equation}
    \mathcal S_i^F \coloneqq \Lambda^\bullet(\mathcal S_i)=\mathcal O\oplus\mathcal S_i,
\end{equation}
so that we can get the Chern character of fermionic Fock sheaf
\begin{equation}
    \mathrm{CH}(\mathcal S_i^F)=1+e^{y_i}=1+e^{-\beta x_i},
\end{equation}
the usual single-level fermionic factor. We now give the schematic diagram for fermionic local Fock sheaf :
\begin{equation}
    \begin{tikzcd}[row sep=small]
\mathcal S_i \arrow[r, "\Lambda^\bullet"] & \mathcal S_i^F \arrow[r, "\mathrm{CH}"] & 1+e^{y_i}.
\end{tikzcd}
\end{equation}
 
To facilitate the following description, we define the fermion sheaf $S^F$ and the bosonic sheaf $S^B$, which are the tensor products of the local Fock Bosonic sheaf and the local fermionic Fock sheaf, respectively: \footnote{In this case, the tensor product is equivalent to the direct product.}
\begin{equation} 
    S^B=\bigotimes_i S^B_i \qquad , \qquad S^F=\bigotimes_i S^F_i.
\end{equation}
Using the splitting principle of Chern character $\mathrm{CH}(E\otimes F)=\mathrm{CH}(E) \cdot \mathrm{CH}(F)$, we can show that the grand partition functions of bosons and fermions are the chern characters of the sheaves $S^B$ and $S^F$, respectively:
\begin{equation} \label{chern character of Bosons}
    \mathrm{CH}\bigl(\operatorname{Sym}^\bullet\mathcal V\bigr)
=\prod_i \mathrm{CH}\bigl(\operatorname{Sym}^\bullet\mathcal S_i\bigr)
=\prod_i \frac{1}{1-e^{y_i}}=\prod_i\frac{1}{1-e^{-\beta x_i}}=\Xi^B,
\end{equation}
\begin{equation} \label{chern character of Fermions}
    \mathrm{CH}\bigl(\Lambda^\bullet\mathcal V\bigr)
=\prod_i \mathrm{CH}\bigl(\Lambda^\bullet\mathcal S_i\bigr)
=\prod_i (1+e^{y_i})=\prod_i\bigl(1+e^{-\beta x_i}\bigr)=\Xi^F.
\end{equation}

\section{The “physical” derivation of the Atiyah-Singer index theorem from non-interacting system}
In this section, we will give a “physical” derivation of the Atiyah-Singer index theorem from non-interacting system. Our so-called “physical” derivation does not refer to starting from the time-dependent evolution of a certain dynamical system, but rather to directly interpreting the topological invariants of the index theorem as the fundamental thermodynamic quantities of non-interacting quantum statistical systems -- the grand partition function. This interpretation based on exact algebraic correspondence provides a novel and profound understanding of the index theorem based on statistical physics.

To facilitate our following analysis, we can rewrite the partition function for the $i$-th single-particle state.    For the Bose-Einstein distribution, \eqref{XiBEi2} can be rewritten as:
\begin{equation} \label{XiBEi3}
\Xi^{BE}_i = \frac{1}{1 - e^{-(\alpha + \beta \epsilon_i)}} = \frac{1}{1 - e^{-x_i}},
\end{equation}
while for the Fermi-Dirac distribution, \eqref{XiFDi1} becomes:
\begin{equation} \label{XiFDi2}
\Xi^{FD}_i = 1 + e^{-\alpha - \beta \epsilon_i} = 1 + e^{-x_i},
\end{equation}
where we have defined \(x_i = \alpha + \beta \epsilon_i\). 

It is important to note that while in the physical context \(x_i\) are real parameters determined by energy level and chemical potential, in our mathematical treatment we will “analytically continue” \(x_i\) to the complex domain. This generalization allows us to make direct contact with characteristic classes in differential geometry, where the variables \(x_i\) correspond to Chern roots -- formal eigenvalues of curvature forms that can take complex values. Therefore, we will treat $x_i$ as a complex variable.

This complex form is not merely a mathematical form. By transforming physical parameters into mathematical variables, we can further utilize the index theorem to rigorously link statistical mechanical distributions with topological invariants.

In fact, in mathematics traditional index theorems consider finite-dimensional vector bundles, while in statistical mechanics we should also consider finite-level fermionic and boson systems. Combining with \eqref{XiBEi3} and \eqref{XiFDi2}, the partition functions are written as follows:
\begin{equation} \label{XiBe N}
    \Xi^{BE}_{N}=\prod^N_i \Xi^{BE}_i=\prod^N_i\frac{1}{1 - e^{-x_i}},
\end{equation}
and 
\begin{equation} \label{XiFd N}
    \Xi^{FD}_{N}=\prod^N_i \Xi^{FD}_i=\prod^N_i(1 + e^{-x_i}),
\end{equation}
where $N$ represents the energy levels of boson systems and fermion systems.

Following the constructions in Section 3, for a finite set of single-particle energy levels ( \(i = 1, 2, ..., N\) ), we obtain the total bosonic and fermionic sheaves on the manifold \(M\):
\begin{equation}
    \mathcal{S}^B_N = \bigotimes_{i=1}^{N} \operatorname{Sym}^{\bullet}(\mathcal{S}_i), \qquad \mathcal{S}^F_N = \bigotimes_{i=1}^{N} \Lambda^{\bullet}(\mathcal{S}_i) = \bigotimes_{i=1}^{N} (\mathcal{O}_M \oplus \mathcal{S}_i),
\end{equation}
where each \(\mathcal{S}_i\) is the line bundle associated with the \(i\)-th energy level.

A fundamental correspondence in differential and algebraic geometry bridges our physical construction to the classical framework of the Atiyah-Singer index theorem: “a locally free sheaf of finite rank on a (complex) manifold is equivalent to a vector bundle”. This principle, rooted in the foundational work of Serre and Swan, establishes a rigorous dictionary between the language of sheaves and the language of fiber bundles \cite{Serre 1955}. 

We now analyze the sheaves \(\mathcal{S}^F_N\) and \(\mathcal{S}^B_N\) under this correspondence:

1.  The Fermionic Sheaf \(\mathcal{S}^F_N\): It is the tensor product of \(N\) locally free sheaves, each of rank $2$ ( \(\mathcal{O}_M \oplus \mathcal{S}_i\) ). Consequently, \(\mathcal{S}^F_N\) itself is a “locally free sheaf of finite rank”, specifically of rank \(2^N\). By the aforementioned equivalence, it corresponds canonically to a smooth (complex) vector bundle over \(M\) of rank \(2^N\).

2.  The Bosonic Sheaf \(\mathcal{S}^B\): Each factor \(\operatorname{Sym}^{\bullet}(\mathcal{S}_i)\) is a free sheaf, but generated by the countably infinite set \(\{\mathcal{O}_M, \mathcal{S}_i, \mathcal{S}_i^{\otimes 2}, \cdots \}\). Therefore, \(\mathcal{S}^B_N\) is a “free sheaf of infinite rank”. While not a finite-rank vector bundle in the classical sense, such infinite-dimensional bundles (or Fock space bundles) are standard objects in geometric quantisation and quantum field theory. Crucially, its Chern character \(\operatorname{ch}(\mathcal{S}^B_N)\) can be rigorously defined via the splitting principle and treated as a formal power series, whose explicit form matches the grand partition function.

Therefore, within the finite-level framework, the Fock sheaves (\(\mathcal{S}^B_N\) and \(\mathcal{S}^F_N\)) constructed from quantum statistics in Section 3 can be interpreted rigorously as vector bundles (finite or infinite-dimensional) on the manifold \(M\). This identification is the pivotal geometric bridge connecting our statistical mechanical formalism to topology.

Based on this, the grand partition functions \(\Xi_N^{BE}\) and \(\Xi_N^{FD}\) in eqs.\eqref{XiBe N} and \eqref{XiFd N} acquire a precise geometric interpretation: they are the explicit forms of the Chern characters of the vector bundles \(\mathcal{S}^B_N\) and \(\mathcal{S}^F_N\), respectively, under the given parametrization. This completes the transition from statistical mechanical objects (partition functions) to differential geometric objects (Chern characters of bundles).

This transition also clarifies the change in the meaning of the parameters \(x_i\): from thermodynamic variables \(x_i = \alpha + \beta \epsilon_i\) in physics, they are naturally reinterpreted as Chern roots of the relevant vector bundles (i.e., eigenvalues of the curvature form) in geometry. This “analytic continuation” or re-interpretation provides the algebraic foundation for the subsequent index theorem equalities.

In summary, by working within the finite-level regime, we embed the physical sheaf structures of Section 3 into the classical framework of vector bundle theory. This grants mathematical legitimacy to the algebraic manipulations in Section 4, where \(\Xi^{BE}_i\) and \(\Xi^{FD}_i\) are directly treated as characteristic classes and linked to the index theorems for Dirac and de Rham operators. One of the core findings of this work -- the profound correspondence between quantum statistical distributions (Bose-Einstein and Fermi-Dirac) and topological invariants (characteristic classes) -- is firmly established upon this correspondence.

Before we start our formal discussion, based on the content of Section 3, we provide the following dictionary of the corresponding mathematical and physical concepts (Table 1).
\begin{table}[ht]
\centering
\caption{Physics-topology dictionary} 
\begin{tabular}{|c|c|}
\hline
$ \textbf{Physical Quantities (Non-Interacting system)} $ & $\textbf{Topological concepts}$ \\
\hline
$\text{Single-particle state} \ $i$ $ & $\text{Line bundle} \ L_i$ \\
\hline
$\text{Energy} \ \epsilon_i $, Chemical potential $\mu$& Parameter $x_i$ (Chern root) \\
\hline
Grand partition function $\Xi$ & Chern character $\operatorname{ch}$\\
\hline
Bosonic Fock space $(n_i=0,1,2,\cdots)$ & Symmetric algebra $\mathrm{Sym}^*$\\
\hline
Fermionic Fock space $n_i=0,1$ & Exterior algebra $\Lambda^*$ \\
\hline
\end{tabular}
\end{table}
\subsection{“Fermi-Bose” pair  for Dirac operator}
Firstly, we assume that \(M\) is a $m=2l=4k$-dimensional compact orientable manifold without a boundary, \(E\) is a spinor bundle on \(M\), and \(D\) is the associated Dirac operator. The index theorem of $D$ is:
\begin{equation} \label{AS for D}
    \text{index}(D) = \int_M \hat{A}(TM) \, \mathrm{ch}(E),
\end{equation}
where \(\hat{A}(TM)\) is the \(\hat{A}\)-class of the tangent bundle, and \(\mathrm{ch}(E)\) is the Chern character of the bundle \(E\).
Let \(x_i\) be the root of \(TM\), then according to \cite{ABP1973}, the definiton of Chern character of spin bundle is
\begin{equation} \label{chE1}
\mathrm{ch}(E)= \prod_i^l (e^{x_i/2}+ e^{-x_i/2}).
\end{equation}
Then combining with \eqref{XiFDi2}, eq.\eqref{chE1} can be rewritten as
\begin{equation} \label{chE2}
\mathrm{ch}(E)= \prod_i^l e^{x_i/2}(1+ e^{-x_i})= \prod_i^l e^{x_i/2} \Xi^{FD}_i.
\end{equation}
The $\hat{A}$-genus is defined by \cite{ABP1973}:
\begin{equation} \label{Ahat}
\hat{A}= \prod_i^l \frac{x_i}{e^{x_i/2}- e^{-x_i/2}}=\prod_i^l \frac{x_i}{e^{x_i/2}(1- e^{-x_i})}.
\end{equation}
Therefore, we have
\begin{equation}\label{chEAhat1}
\begin{split}
\mathrm{ch}(E) \cdot \hat{A} &= \prod_i^le^{x_i/2} \Xi^{FD}_i \cdot \prod_i^l \frac{x_i}{e^{x_i/2}(1- e^{-x_i})}\\
          &= \prod_i^l e^{x_i/2}(1+ e^{-x_i}) \frac{x_i}{e^{x_i/2}(1- e^{-x_i})}\\
          &= \prod_i^l \frac{1+ e^{-x_i}}{1- e^{-x_i}} \prod_i^l x_i \\
          &= \prod_i^l (\Xi^{BE}_i \Xi^{FD}_i) \cdot \prod_i^l x_i.
\end{split}
\end{equation}
We can notice that in \eqref{XiBEi2} there is a minus sign for Bose-Einstein distribution and in \eqref{XiFDi1} there is a plus sign for Fermi-Dirac distribution. In \eqref{chE1}, there is a plus sign in the product. In \eqref{Ahat}, there is a minus sign in the product. Therefore, if there is plus sign in the product, we call it the “Fermi” product while if there is minus sign in the product, we call it the “Bose” product. So, for \eqref{chEAhat1}, we can call it the “Fermi-Bose” pair.

Furthermore,we extend the non-degenerate condition $e^{\alpha} \gg 1$ or $e^{-\alpha} \ll 1$ in physics to $e^{x_i} \gg 1$ or $e^{-x_i} \ll 1$, and apply it to subsequent calculations. \eqref{chEAhat1} becomes
\begin{equation} \label{chEAhat2}
    \mathrm{ch}(E) \cdot \hat{A} \approx \prod_i^l x_i,
\end{equation}
and the index is given by:
\begin{equation} \label{ind1}
    \text{index}= \int_M \mathrm{ch}(E) \cdot \hat{A} \approx \int_M\prod_i^l x_i = \chi(M),
\end{equation}
where $\chi(M)$ is the Euler class. In other words, the classical non-interacting system can give the Euler class.

\subsection{“Bose-Bose” pair for de Rham operator}
Let $M$ be an $m$-dimensional compact orientable manifold with no boundary and  $M$ is even dimensional.
In this subsection, we consider \(D_R\) as a de Rham operator, whose analytic index is 
\begin{equation} \label{index for DR}
    \text{index}(D_R) = \chi(M),
\end{equation}
where $\chi(M)$ is the Euler characteristic of $M$ \cite{nakahara2018geometry}. Next, we will consider topological index:
\begin{equation} \label{toplogical index}
    (-1)^{l(2l+1)}\int_M \operatorname{ch}\left(\bigoplus_{r=0}^m(-1)^r \bigwedge^r T^*M^{\mathbb{C}} \right)\frac{\mathrm{Td}(TM^{\mathbb{C}})}{e(TM)}
\end{equation}
where $m=2l$, $T^*M^{\mathbb{C}}$ is the dual space of $TM^{\mathbb{C}}$ and
\begin{equation} \label{eTM}
    e{(TM)}=\prod_i^l x_i(TM^{\mathbb{C}}).
\end{equation}

Let \(x_i\) be a Chern root of \(TM^\mathbb{C}\), then the Euler class \(e(TM) = \prod_i x_i\).

In formal calculation, the Chern character of symbol bundle of the de Rham complex \(\sigma(D)=\bigoplus_{r=0}^m(-1)^r \bigwedge^r T^*M^{\mathbb{C}}\) is rewitten as:
\begin{equation} \label{Chern character of sigma D}
    \operatorname{ch}(\sigma(D)) = \prod_i^m (1 - e^{-x_i})(TM^{\mathbb{C}}),
\end{equation}
Combining with \eqref{XiBEi3}, eq.\eqref{Chern character of sigma D} becomes
\begin{equation} \label{sigma D and BE}
    \operatorname{ch}(\sigma(D))= \prod_i^m\frac{1}{\Xi_i^{BE}},
\end{equation}
and the Todd class is:
\begin{equation} \label{td}
    \operatorname{Td}(TM^{\mathbb{C}}) = \prod_i^m \frac{x_i}{1 - e^{-x_i}}=\prod_i^m\Xi_i^{BE} \prod_i^mx_i.
\end{equation}
Therefore, the topological index is:
\begin{equation}
    \operatorname{index}= \left(\prod_i^m\frac{1}{\Xi_i^{BE}} \right)\frac{\prod_i^m\Xi_i^{BE} \prod_i^mx_i}{e(TM)}=\chi(M).
\end{equation}
Therefore, we can also obtain the Euler class from the de Rham operator. This is consistent with the index theorem. 

\subsection{“Fermi-Fermi” pair, Dirac operator and $\hat{B}$-genus}
In this subsection, we introduce a new characteristic class -- the “$\hat{B}$-genus” -- motivated by the Fermi-Dirac distribution, and demonstrate its role in the index theorem for the Dirac operator.\\
\textbf{Mathematical motivation for the $\hat{B}$-genus}

Inspired by the duality between statistical distributions and characteristic classes ($\hat{A}$-genus \eqref{Ahat}), we define the $\hat{B}$-genus using a similar but different generating function:
\begin{equation} \label{B class}
    \hat{B}(M) = \prod_{j=1}^{l} \frac{x_j}{e^{x_j/2} + e^{-x_j/2}}.
\end{equation}
This definition naturally arises from the Fermi-Dirac distribution and establishes a “quasi characteristic class” to the \(\hat{A}\)-genus, much as hyperbolic cosine complements hyperbolic sine.\\
\textbf{Index theorem from $\hat{B}$-genus}

For the Dirac operator \(D\) acting on sections of a spin bundle \(E\) over a spin manifold \(M\), and using the $\hat{B}$-genus, we obtain an equivalent but novel formulation to describe the topological index:
\begin{equation}
    \text{index}(D) = \int_M \mathrm{ch}(E) \cdot \hat{B}(TM).
\end{equation}
Combining with \eqref{B class} and \eqref{chE1}, we can get
\begin{equation}
    \int_M \mathrm{ch}(E) \cdot \hat{B}(TM) = \int_M\prod_i^l x_i=\chi(M).
\end{equation}
In other words, we can obtain the Euler class from $\hat{B}$-genus.

This result reveals a profound connection, that is, the index of the Dirac operator, when expressed via the $\hat{B}$-genus, yields the Euler class. This provides a “quasi” perspective on the topology of the Dirac operator, linking it directly to the topological invariant of the de Rham complex.\\
\textbf{Physical interpretation}

The “Fermi-Fermi” pairing reflects the purely fermionic nature of the Dirac operator. In our analogy of statistical mechanics, we have:\\
- The spin bundle \(E\) corresponds to the fermionic Fock space;\\
- The $\hat{B}$-genus embodies the exclusion principle of fermions in the context of characteristic classes.\\
This pairing demonstrates that the statistical distribution of fermions alone can capture deep topological invariants without relying on supersymmetry.

\subsection{“Bose-Fermi” pair for de Rham operator}
Similar to Section 4.3, we can define a quasi Todd class:
\begin{equation} \label{dual Todd class}
    \operatorname{Td}^*(TM^{\mathbb{C}})= \prod_i^m \frac{x_i}{1 + e^{-x_i}}=\prod_i^m\frac{1}{\Xi_i^{FD}} \prod_i^mx_i.
\end{equation}
The index theorem for the de Rham operator is:
\begin{equation} \label{DR2}
     \operatorname{index}=\int_M \operatorname{ch}(\sigma(D))\frac{\operatorname{Td^*}(TM^{\mathbb{C}})}{e(TM)}.
\end{equation}
Combining with \eqref{Chern character of sigma D} and \eqref{sigma D and BE}, we can get:
\begin{equation}
\begin{split}
     \operatorname{index}&=\int_M \prod_i^m (1 - e^{-x_i})(TM^{\mathbb{C}})\prod_i^m \frac{x_i}{1 + e^{-x_i}} \frac{1}{e(TM)}\\
     &=\int_M \left(\prod_i^m\frac{1}{\Xi_i^{BE}} \right)\prod_i^m\frac{1}{\Xi_i^{FD}} \prod_i^lx_i\\
     &=\int_M \prod_i^m\frac{1 - e^{-x_i}}{1 + e^{-x_i}}\prod_i^lx_i
\end{split}
\end{equation}
Furthermore, we extend the non-degenerate condition $e^{\alpha} \gg 1$ or $e^{-\alpha} \ll 1$ in physics to $e^{x_i} \gg 1$ or $e^{-x_i} \ll 1$, and apply it to subsequent calculations. The index becomes
\begin{equation}
    \operatorname{index} \approx \int_M \prod_i^lx_i=\chi(M)
\end{equation}
This result is consistent with the result for the de Rham operator in subsection 4.2.\\

Tables 2 and 3 illustrate the relationships among four different pairs, comparing the corresponding operators, vector bundles, Chern characters, genus, index theorems, and topological invariants.

In this section, we have derived the Atiyah–Singer index theorem directly from non-supersymmetric quantum statistics by establishing a fundamental correspondence: the grand partition functions of non-interacting bosonic and fermionic systems coincide precisely with the Chern characters of certain vector bundles. Through four distinct pairings -- Fermi–Bose, Bose–Bose, Fermi–Fermi, and Bose–Fermi -- we recovered the index formulas for both the Dirac and de Rham operators, while naturally introducing two new characteristic classes, the \(\hat{B}\)-genus and the \(\mathrm{Td}^*\)-class, which dualize the classical \(\hat{A}\)-genus and Todd class. This work demonstrates that the deep topology captured by the index theorem can emerge purely from quantum statistical mechanics, without reliance on supersymmetry, thereby offering a fresh and unifying perspective on the interplay between statistical physics and differential topology.
\begin{table}[ht]
\centering
\caption{Dictionary of Fermi-Bose pair and Bose-Bose pair} 
\begin{tabular}{|c|c|c|}
\hline
$ \textbf{Pairing} $ & $\text{Fermi-Bose}$ & $\text{Bose-Bose}$ \\
\hline
$\textbf{Operator}  $ & $\text{Dirac}$ & $\text{de Rham}$\\
\hline
$\textbf{Vector bundle} $ & $\text{Spin bundle}\ E$ & $\text{de Rham complex symbol bundle}\ \sigma(D)$ \\
\hline
$\textbf{Chern character}$ & $\prod_i^m (e^{x_i/2}+e^{-x_i/2})=\prod_i^m e^{x_i/2} \Xi^{FD}_i$ & $\prod_i^m (1 - e^{-x_i})=\prod_i^m\frac{1}{\Xi_i^{BE}}$\\
\hline
$\textbf{Topological class}$ &$\hat{A}-\text{genus}$ &Todd class $\operatorname{Td}$ \\
\hline
$\textbf{Index theorem}$ & $\int_M \hat{A}(TM)\mathrm{ch}(E)$ & $\int_M \mathrm{ch}(\sigma(D))\frac{\mathrm{Td}(TM)}{e(TM)}$ \\
\hline
\end{tabular}
\end{table}
\begin{table}[ht]
\centering
\caption{Dictionary of Fermi-Fermi pair and Bose-Fermi pair} 
\begin{tabular}{|c|c|c|}
\hline
$ \textbf{Pairing} $ & $\text{Fermi-Fermi}$ & $\text{Bose-Fermi}$ \\
\hline
$\textbf{Operator}  $ & $\text{Dirac}$ & $\text{de Rham}$\\
\hline
$\textbf{Vector bundle} $ & $\text{Spin bundle}\ E$ & $\text{de Rham complex symbol bundle}\ \sigma(D)$ \\
\hline
$\textbf{Chern character}$ & $\prod_i (e^{x_i/2}+e^{-x_i/2})=\prod_i^m e^{x_i/2} \Xi^{FD}_i$ & $\prod_i^m (1 - e^{-x_i})=\prod_i^m\frac{1}{\Xi_i^{BE}}$\\
\hline
$\textbf{Topological class}$ &$\hat{B}-\text{genus}$ &quasi Todd class $\operatorname{Td^*}$ \\
\hline
$\textbf{Index theorem}$ & $\int_M \hat{B}(TM)\mathrm{ch}(E)$ & $\int_M \mathrm{ch}(\sigma(D))\frac{\mathrm{Td}^*(TM)}{e(TM)}$ \\
\hline
\end{tabular}
\end{table}

\section{Infinite-dimensional Case of Traditional Index Theorem}
In Sections 3 and 4, we established a concrete correspondence between the partition functions of quantum statistical systems and the characteristic classes of finite-dimensional vector bundles, leading to novel derivations of the Atiyah-Singer index theorem. This chapter aims to generalize the correspondence into a more abstract and general mathematical framework. We propose that the deep link between quantum statistics and topology can be axiomatized by starting from the spectral data of a suitable operator acting on a sheaf. This allows us to define topological quantities like Euler numbers and genera directly from this spectral data, independent of any pre-existing geometric structure like a connection or curvature. The goal is to construct an abstract index theorem within this new framework, where the objects are sheaf-endomorphism pairs. However, we should pointed out that the promotion of this chapter is merely a formalistic attempt.

\subsection{From spectrum to geometry}
In classical index theory, the Chern character of a vector bundle is defined via curvature forms. For sheaves equipped with an operator structure, we can directly define the Chern character from the spectrum of the operator by interpreting eigenvalues as Chern roots. To provide a rigorous geometric foundation for this construction, we introduce the infinite-dimensional complex projective space $\mathbb{CP}^{\infty}$ as a classifying space, identifying each eigenvalue with the first Chern class of a line bundle. This section elaborates on this construction and demonstrates its compatibility with the classical theory of characteristic classes.

Let \(M\) be a compact smooth manifold without boundary, and let \(\mathcal{S}\) be a complex vector bundle (or a locally free sheaf) over \(M\). Consider a self-adjoint differential operator \(F: \Gamma(M, \mathcal{S}) \to \Gamma(M, \mathcal{S})\) with a discrete spectrum \(\{\lambda_i\}_{i=1}^\infty\). We assume that the generalized eigenspace corresponding to each eigenvalue \(\lambda_i\) forms a smooth subbundle \(\mathcal{S}_i \subset \mathcal{S}\), leading to a direct sum decomposition:
\begin{equation}
    \mathcal{S} = \bigoplus_{i=1}^\infty \mathcal{S}_i,
\end{equation}
such that \(F|_{\mathcal{S}_i} = \lambda_i \cdot \mathrm{id}_{\mathcal{S}_i}\). If an eigenvalue has multiplicity, we can relabel the indices so that each distinct eigenvalue (accounting for multiplicity) corresponds to a one-dimensional subbundle. Without loss of generality, we therefore assume that each \(\mathcal{S}_i\) has rank one, i.e., each eigenvalue corresponds to a line bundle.

To connect the spectrum of the operator \(F\) with characteristic classes in differential geometry, we need to lift each real eigenvalue \(\lambda_i\) to a closed 2-form. Assume there exist closed 2-forms \(\omega_i \in \Omega^2(M)\) (with \(d\omega_i = 0\)) that are integral (their periods lie in \(2\pi i\mathbb{Z}\)) and satisfy:
\begin{equation}
    \frac{i}{2\pi} \omega_i = \lambda_i \quad \text{(as constant functions)}.
\end{equation}
More precisely, we require that the cohomology class \([\omega_i] \in H^2(M; 2\pi i\mathbb{Z})\). Through geometric quantization, such a class corresponds to a complex line bundle \(L_i \to M\) with first Chern class \(c_1(L_i) = [\omega_i]\).

To handle all eigenvalues uniformly, we introduce the infinite-dimensional complex projective space \(\mathbb{CP}^\infty\). As a classifying space, the universal line bundle \(\mathcal{O}(1)\) over \(\mathbb{CP}^\infty\) satisfies: for any compact manifold \(M\), isomorphism classes of complex line bundles over \(M\) correspond bijectively to homotopy classes of maps \([M, \mathbb{CP}^\infty]\), and \(c_1(\mathcal{O}(1)) = H\) (the hyperplane class). For each eigenvalue \(\lambda_i\), we construct a classifying map \(f_i: M \to \mathbb{CP}^\infty\) such that the pullback bundle satisfies:
\begin{equation}
    L_i = f_i^* \mathcal{O}(1), \quad c_1(L_i) = [\omega_i]=-\lambda.
\end{equation}
In this way, each eigenvalue \(\lambda_i\) of the operator is geometrically realized as a line bundle \(L_i\). The geometric information of the entire spectral sheaf pair \((\mathcal{S}, F)\) can thus be encoded by a family of maps \(f_i: M \to \mathbb{CP}^\infty\).
For details, we can refer to the operation of the Hamiltonian operator in \cite{Li:2025qry}.

With this geometric realization, we can define the Chern character of each eigen-subbundle \(\mathcal{S}_i\) as:
\begin{equation}
    \operatorname{ch}(\mathcal{S}_i) := \operatorname{ch}(L_i) = e^{[\omega_i]}=e^{-\lambda_i},
\end{equation}
where \(e^{\omega_i} = 1 + \omega_i + \frac{\omega_i^2}{2!} + \cdots\) is the formal exponential series. Here we assume that the curvature form of \(L_i\) is chosen to be \(\omega_i\), so that \(\frac{i}{2\pi} F_{L_i} = \omega_i\). The Chern character of the entire bundle \(\mathcal{S}\) is then defined as:
\begin{equation}
    \operatorname{ch}(\mathcal{S}) := \sum_{i=1}^{\infty} \operatorname{ch}(\mathcal{S}_i) = \sum_{i=1}^{\infty} e^{-\lambda_i}.
\end{equation}
Under suitable convergence conditions (for instance, if the eigenvalues grow sufficiently fast, or if we work with formal power series), this expression is well-defined as a differential form.

When the operator \(F\) happens to be the curvature operator of some Hermitian connection \(\nabla\), i.e., \(F = \frac{i}{2\pi} F_\nabla\), then the eigenvalues of the curvature form \(F_\nabla\) are themselves closed 2-forms. In this case, our defined Chern character coincides with the classical Chern character:
\begin{equation}
    \operatorname{ch}(\mathcal{S}) = \operatorname{Tr} e^{\frac{i}{2\pi} F_\nabla} = \sum_{i} e^{-\lambda_i}.
\end{equation}
Thus, our construction can be viewed as a natural generalization of the classical Chern character in the language of spectral geometry. This purely geometric framework, which directly links the spectral data of an operator to characteristic classes via the $\mathbb{CP}^{\infty}$ model, provides a foundation for studying index theorems on arbitrary manifolds.
\subsection{The abstract category of spectral sheaves}
We will continue with the convention from the previous subsection:\\
Let \( M \) be a compact manifold. The fundamental objects of our theory are pairs \( (\mathcal{S}, F) \), where:\\
1. \( \mathcal{S} \) is a sheaf over \( M \). For physical and geometric concreteness, we typically consider \( \mathcal{S} \) to be a locally free sheaf of finite rank (i.e., a vector bundle) or its natural infinite-rank generalizations (e.g., complete symmetric algebras).\\
2. \( F: \mathcal{S} \to \mathcal{S} \) is an endomorphism (a sheaf morphism). To endow the spectrum with topological meaning, we require \( F \) to possess a well-defined, discrete spectrum under appropriate analytical conditions. Operationally, we assume that it is possible to associate to \( F \) a countable set of formal eigenvalues \( \{\lambda_i\} \). In practice, \( F \) should possess a zeta-function regularizable determinant.

The core physical data of the pair \( (\mathcal{S}, F) \) is the spectrum \( \{\lambda_i\} \) of the operator \( F \), which in a quantum system corresponds to the single-particle energy levels. We have already witnessed an elegant correspondence: the exponential sum of the spectrum, \( \sum_i e^{-\lambda_i} \), gives the grand partition function of the system, and this partition function is mathematically identified as the Chern character of a sheaf (or vector bundle).

This correspondence leads to a natural and profound question: if a sum (in this exponential form) of the spectrum encodes one topological invariant (the Chern character), then what should the product of the spectrum, \( \prod_i \lambda_i \) -- the most direct combinatorial invariant measuring the scale of the operator -- correspond to?

In differential geometry, the product of all non-zero eigenvalues of a natural operator (e.g., the Laplacian) on a manifold, once regularized, is intimately related to the manifold's Euler characteristic. More precisely, the product of the eigenvalues of the curvature form (the Chern roots) defines the Euler class. This is no coincidence but reveals a universal source of topological invariants: they can often be expressed as the regularized determinant of a natural operator's spectrum.

Therefore, for our spectral-sheaf pair \( (\mathcal{S}, F) \), we identify the (formal) product of its eigenvalues as an “formal Euler class” for the system. We denote it as:

\begin{equation} \label{formal Euler class}
    \chi(\mathcal{S}, F) := \int_M\prod_i \lambda_i.
\end{equation}

As a formal expression, this product may be divergent. To extract a finite topological invariant from it, a regularization scheme is required, such as zeta-function regularization, which converts this formal product into a well-defined number, the regularized determinant \( \operatorname{det}_{\zeta}'(F) \):
\begin{equation} \label{zeta regularation}
    \operatorname{det}_{\zeta}'(F) := \exp\left(-\frac{d}{ds}\zeta_F(s)\big|_{s=0}\right),
\end{equation}
where \( \zeta_F(s) = \sum_i x_i^{-s} \) is the spectral zeta function of \( F \). 

\subsection{The generalized index theorem}
With the basic object of the spectral-sheaf pair \( (\mathcal{S}, F) \) and its associated abstract Euler class defined, we can now construct a complete algebraic structure that ultimately leads to a generalized index theorem. This construction process is essentially an attempt to abstract and elevate the correspondence established in Section 4 into a universal mathematical framework.

Firstly, starting with the spectral pairs \( (\mathcal{S}, F) \) and their spectrum \( \{\lambda_i\} \), we use standard algebraic operations to represent the construction of the Fock space of quantum systems. The construction of symmetric algebras and external algebras naturally reflects the algebraic structures of Bose statistics and Fermi statistics, respectively:

1.  Bosonic (Symmetric) Construction: we form the symmetric algebra sheaf \( \operatorname{Sym}^\bullet (\mathcal{S}) = \bigoplus_{n \geq 0} \operatorname{Sym}^n (\mathcal{S}) \). Its Chern character, determined by the spectrum \( \{\lambda_i\} \), takes the form of an infinite product:
    \begin{equation} \label{XiBSF}
         \Xi_{BE}(\mathcal{S}, F) := \prod_i \frac{1}{1 - e^{-\lambda_i}}.
    \end{equation}
  Just as \eqref{chern character of Bosons}, this is precisely the grand partition function of an ideal Bose gas with a single-particle spectrum \( \{\lambda_i\} \) . In Section 4, we saw that this form coincides with part of the Chern character and the Todd class of a vector bundle.

2.  Fermionic (Exterior) Construction: we form the exterior algebra sheaf \( \Lambda^\bullet (\mathcal{S}) = \bigoplus_{n \geq 0} \Lambda^n (\mathcal{S}) \). Its corresponding formal character is:
\begin{equation} \label{xiFSF}
    \Xi_{FD}(\mathcal{S}, F) := \prod_i (1 + e^{-\lambda_i}).
\end{equation}
  Just as \eqref{chern character of Fermions}, this also corresponds to the grand partition function of an ideal Fermi gas and is linked in geometry to the Chern character of the exterior algebra and objects like the \( \hat{A} \)-genus.

 These formal products generated directly from the spectrum provide us with the fundamental building blocks. Our aim is to synthesize them into a Generalized Index Theorem of the following form: for a spectral-sheaf pair \( (\mathcal{S}, F) \) suitably linked to an elliptic complex, there exists a topological index density \(\mathcal{I}(\mathcal{S}, F)\) such that
\begin{equation} \label{general index TH}
    \operatorname{index}(D) = \int_M \mathcal{I}(\mathcal{S}, F),
\end{equation}
where \(\mathcal{I}(\mathcal{S}, F)\) is a specific combination of the formal products associated with \((\mathcal{S}, F)\).

We need to consider the four pairings (Fermi-Bose, Fermi-Fermi, Bose-Bose, Bose-Fermi) in Section 4 under the spectral-sheaf pair \( (\mathcal{S}, F) \), but before that, we need to extend the two finite-dimensional vector bundles (spinor bundle $E$ and de Rham complex symbol bundle $\sigma(D)$) to infinite dimensions.

In the classical finite-dimensional theory, the constructions of the spinor bundle and the de Rham complex symbol bundle explicitly depend on the manifold's dimensions \(l\) and \(m=2l\). To reconstruct this structure within the infinite-dimensional framework of the spectral-sheaf pair \( (\mathcal{S}, F) \), we must find an intrinsic rationale to replace the concept of finite dimension. This rationale is naturally provided by the operator's spectral symmetry.

We introduce a key structure for the spectral-sheaf pair \( (\mathcal{S}, F) \): a sheaf endomorphism \(\Gamma: \mathcal{S} \to \mathcal{S}\), called a grading involution. It satisfies \(\Gamma^2 = \text{id}_{\mathcal{S}}\) and maintains a specific (anti-) commutation relation with the original operator \(F\). This operator abstractly corresponds to the chiral operator in classical geometry (such as the Clifford volume element or the Hodge star operator) \cite{HD 2005}.

Under the action of \(\Gamma\), the sheaf \(\mathcal{S}\) decomposes into positive and negative eigen-sheaves:
\begin{equation} \label{sheaf decomposition}
    \mathcal{S} = \mathcal{S}_+ \oplus \mathcal{S}_- , \quad \text{where } \Gamma|_{\mathcal{S}_{\pm}} = \pm \text{id}.
\end{equation}
The (anti-)commutation relation between \(F\) and \(\Gamma\) ensures that \(F\) (or its square) preserves \(\mathcal{S}_+\) and \(\mathcal{S}_-\) respectively. We denote the (formal) spectrum of \(F\) restricted to \(\mathcal{S}_+\) as \(\{\lambda^+_i\}_{i \in I_+}\) and to \(\mathcal{S}_-\) as \(\{\lambda^-_j\}_{j \in I_-}\). These two spectral sets together constitute the complete spectral data of \(F\), embodying its internal dual structure.

Based on this spectral decomposition, we define two core sheaf constructions corresponding to the infinite-dimensional generalizations:

1.  (Infinite-Dimensional) Spinor-Type Sheaf: this construction mimics the algebraic properties of finite-dimensional spinor bundles, with its character generated by the paired spectrum \(\lambda^+_i\). We define its formal Chern character as:
\begin{equation} \label{new E}
    \mathcal{CH}(\mathcal{S}, F,E) := \prod_{i \in I_+} (e^{\lambda^+_i/2} + e^{-\lambda^+_i/2})
\end{equation}
    Physically, this corresponds to the grand partition function of a fermionic system generated by positive chirality.

2.  (Infinite-Dimensional) de Rham-Type Sheaf: this construction mimics the finite-dimensional de Rham complex symbol bundle, and its character should involve all spectral modes. We define its formal Chern character as:
\begin{equation} \label{new sigma D}
    \mathcal{CH}(\mathcal{S}, F,\sigma(D)) := \prod_{i \in I_+} (1 - e^{-\lambda^+_i}) \cdot \prod_{j \in I_-} (1 - e^{-\lambda^-_j}).
\end{equation}
    This corresponds to the grand partition function of a bosonic system generated by all positive and negative modes. In particular, in the important geometric context of a complex manifold (notably, a Kähler manifold), the spectral symmetry enforced by the complex structure -- via the action of the Hodge star operator and Serre duality -- leads to a perfect pairing between the positive and negative chirality sectors. This results in the identification of their spectra: \(\lambda_i^+ =\lambda_i^-\) for all \(i \in I_+ \cong I_-\). Consequently, the de Rham-type construction simplifies to a perfect square form \(\prod_i (1 - e^{-\lambda_i})^2\), which makes the origin of the classical dimensional relation \(m = 2l\) manifest at the spectral level.

   This spectral symmetry is not merely a special case, but a guiding principle for our generalization. When our abstract spectral-sheaf pair \( (\mathcal{S}, F) \) is concretely realized on a complex manifold -- for instance, by taking \( \mathcal{S} \) to be the sheaf of differential forms and \( F \) a natural operator like the Dolbeault Laplacian -- the grading involution \(\Gamma\) is precisely realized by (a power of) the complex Hodge star operator. Consequently, the defining property \(\lambda_i^+ = \lambda_i^-\) becomes an inherent feature of our framework. Therefore, a consistent and geometrically meaningful realization of our abstract \( (\mathcal{S}, F) \) in the context of complex geometry should and does satisfy this spectral identification, ensuring that the infinite-dimensional de Rham-type construction specializes correctly to its well-studied finite-dimensional counterpart.

Thus, similar to \eqref{chE2} and \eqref{sigma D and BE}, we can rewrite \eqref{new E} and \eqref{new sigma D} as
\begin{equation} \label{R new E}
    \mathcal{CH}(S,F,E)=\prod_{i} (e^{\lambda_i/2} + e^{-\lambda_i/2})=\Xi_{FD}(S,F)\prod_{i}e^{\lambda_i/2}
\end{equation}
and
\begin{equation} \label{R new sigma D}
    \mathcal{CH}(S,F,\sigma(D))=\left(\prod_i(1-e^{-\lambda_i})\right)^2=\left(\frac{1}{\Xi_{BE}(S,F)}\right)^2
\end{equation}

We now use the above construction to rewrite the topological quantities ($\hat{A}$-genus, $\hat{B}$-genus, Todd class and Todd* class) involved in Section 4.\\
1. Formal $\hat{A}$-genus
\begin{equation} \label{formal A}
    \hat{A}(S,F)=\prod_i \frac{\lambda_i}{e^{\lambda_i/2}- e^{-\lambda_i/2}}=\prod_i \frac{\lambda_i}{e^{\lambda_i/2}(1- e^{-\lambda_i})}=\Xi_{BE}(S,F)\prod_i\frac{\lambda_i}{e^{\lambda_i/2}}.
\end{equation}
2. Formal $\hat{B}$-genus
\begin{equation} \label{formal B}
    \hat{B}(S,F)=\prod_i \frac{\lambda_i}{e^{\lambda_j/2} + e^{-\lambda_j/2}}=\prod_i \frac{\lambda_i}{e^{\lambda_i/2}(1+e^{-\lambda_i})}=\frac{1}{\Xi_{FD}(S,F)}\prod_i\frac{\lambda_i}{e^{\lambda_i/2}}.
\end{equation}
3. Formal Todd class
\begin{equation} \label{formal Td}
    \mathrm{Td}(S,F)=\left(\prod_i\frac{\lambda_i}{1-e^{-\lambda_i}}\right)^2=\left(\Xi_{BE}\prod_i\lambda_i\right)^2
\end{equation}
4. Formal Todd* class
\begin{equation} \label{formal Td*}
    \mathrm{Td}^*(S,F)=\left(\prod_i\frac{\lambda_i}{1+e^{-\lambda_i}}\right)^2=\left(\frac{\prod_i\lambda_i}{\Xi_{FD}}\right)^2
\end{equation}

Finally, similar to Section 4, we present the index density for the four pairs:\\
1. Fermi-Bose pair:
\begin{equation}\label{FB pair}
    \mathcal{I}_{FB}=\mathcal{CH}(S,F,E) \cdot \hat{A}(S,F)=\left(\Xi_{FD}(S,F)\prod_{i}e^{\lambda_i} \right) \left(\Xi_{BE}(S,F)\prod_i\frac{\lambda_i}{e^{\lambda_i}}\right).
\end{equation}
We continue applying the extended nondegenerate conditions ($e^{x_i} \gg 1$ or $e^{-x_i} \ll 1$) from Section 4, then \eqref{FB pair} becomes:
\begin{equation} \label{nd FB}
    \mathcal{I}_{FB}=\prod_i\lambda_i.
\end{equation}
After integration, we can obtain the formal Euler class \eqref{formal Euler class}.\\
2. Bose-Bose pair:
\begin{equation}\label{BB pair}
\begin{split}
    \mathcal{I}_{FB}&=\mathcal{CH}(S,F,\sigma(D)) \cdot \frac{\mathrm{Td}(S,F)}{\prod_i\lambda_i}\\
    &=\left(\frac{1}{\Xi_{BE}(S,F)}\right)^2 \left(\Xi_{BE}\prod_i\lambda_i \right)^2 \left(\frac{1}{\prod_i\lambda_i} \right)\\
    &=\prod_i\lambda_i.
\end{split}
\end{equation}
After integration, we can also obtain the formal Euler class.\\
3. Fermi-Fermi pair:
\begin{equation} \label{FF pair}
    \mathcal{I}_{FF}=\mathcal{CH}(S,F,E)\cdot\hat{B}(S,F)=\left(\Xi_{FD}(S,F)\prod_{i}e^{\lambda_i/2}\right) \left(\frac{1}{\Xi_{FD}(S,F)}\prod_i\frac{\lambda_i}{e^{\lambda_i/2}} \right)=\prod_i \lambda_i.
\end{equation}
After integration, we can get the formal Euler class.\\
4. Bose-Fermi pair:
\begin{equation} \label{BF pair}
    \begin{split}
        \mathcal{I}_{BF}&=\mathcal{CH}(S,F,\sigma(D)) \frac{\mathrm{Td}^*(S,F)}{\prod_i\lambda_i}\\
        &=\left(\frac{1}{\Xi_{BE}(S,F)}\right)^2\left(\frac{\prod_i\lambda_i}{\Xi_{FD}(S,F)}\right)^2\frac{1}{\prod_i\lambda_i}\\
        &=\left(\frac{1}{\Xi_{BE}(S,F)\Xi_{FD}(S,F)} \right)^2\prod_i\lambda_i.
    \end{split}
\end{equation}
Similar to the Fermi-Bose pair case, applying the extended nondegenerate conditions ($e^{x_i} \gg 1$ or $e^{-x_i} \ll 1$), we can obtain $\prod_i\lambda_i$, and then integrating on the manifold $M$, we can get $\chi(S,F)$.

In Section 5, the correspondence between quantum statistics and topology is extended to the infinite‑dimensional setting by introducing the central notion of spectral‑sheaf pairs \((\mathcal{S}, F)\). The spectrum \(\{\lambda_i\}\) of the operator \(F\) is geometrically realized as line bundles, allowing the Chern character and the formal Euler class to be defined directly from the spectral data. Within this framework, bosonic and fermionic Fock sheaves are constructed via symmetric and exterior algebras, whose grand partition functions yield \(\Xi_{BE}\) and \(\Xi_{FD}\) respectively. By introducing a grading involution \(\Gamma\), infinite‑dimensional analogues of spinor sheaves and de Rham‑type sheaves are defined. The resulting generalized index theorem shows that, in the non‑degenerate limit, all four pairings reduce to the same formal Euler class \(\prod_i \lambda_i\). This approach offers an intrinsic, spectral‑geometric perspective on topological invariants, going beyond the traditional finite‑dimensional and supersymmetric frameworks.

We provide a dictionary to compare the concepts of traditional geometry with the concepts of infinite-dimensional geometry introduced in Section 5 (Table 4). We should note that most of the concepts introduced in Section 5 require $\zeta$-regularization.

\begin{table}[ht]
\centering
\caption{Dictionary of geometry concepts for finite and infinite dimensions} 
\begin{tabular}{|c|c|}
\hline
traditional geometry concepts & infinite geometry concepts \\
\hline
vector bundle & sheaf $S$ \\
\hline
curvature form $R$ & operator $F$\\
\hline
Chern root $x_i$ & eigenvalues $\lambda_i$ \\
\hline
Euler class $\chi(M)=\prod_i^lx_i$ & formal Euler class $\chi(S,F)=\prod_i\lambda_i$\\
\hline
Chern character $\mathrm{ch}=\mathrm{Tr}(\mathrm{e}^{\frac{i}{2\pi}R})$ & formal Chern character $\mathcal{CH}=\mathrm{Tr}(\mathrm{e}^{-F})$ \\
\hline
$\mathrm{ch}(\sigma(D))=\prod_i^m(1-\mathrm{e}^{-x_i})$&$\mathcal{CH}(S,F,\sigma(D))=(\prod_i(1-\mathrm{e}^{-\lambda_i}))^2$\\
\hline
$\mathrm{ch}(E)=\prod_i^l(\mathrm{e}^{-\frac{x_i}{2}}+\mathrm{e}^{\frac{x_i}{2}})$ & $\mathcal{CH}(S,F,E)=\prod_i(\mathrm{e}^{-\frac{\lambda_i}{2}}+\mathrm{e}^{\frac{\lambda_i}{2}})$\\
\hline
$\hat{A}(\mathrm{TM})=\prod_i^l\frac{x_i}{\mathrm{e}^{x_i/2}-\mathrm{e}^{-x_i/2}}$ & $\hat{A}(S,F)=\prod_i\frac{\lambda_i}{\mathrm{e}^{\lambda_i/2}-\mathrm{e}^{-\lambda_i/2}}$\\
\hline
$\hat{B}(\mathrm{TM})=\prod_i^l\frac{x_i}{\mathrm{e}^{x_i/2}+\mathrm{e}^{-x_i/2}}$ & $\hat{B}(S,F)=\prod_i\frac{\lambda_i}{\mathrm{e}^{\lambda_i/2}+\mathrm{e}^{-\lambda_i/2}}$\\
\hline
 $\mathrm{Td}(\mathrm{TM})=\left(\prod_i^l\frac{x_i}{1-e^{-x_i}}\right)^2$ &  $\mathrm{Td}(S,F)=\left(\prod_i\frac{\lambda_i}{1-e^{-\lambda_i}}\right)^2$\\
 \hline
$\mathrm{Td}^*(\mathrm{TM})=\left(\prod_i^l\frac{x_i}{1+e^{-x_i}}\right)^2$ & $\mathrm{Td}^*(S,F)=\left(\prod_i\frac{\lambda_i}{1+e^{-\lambda_i}}\right)^2$\\
\hline
$I_{FB}=\int_M\hat{A}(\mathrm{TM})\mathrm{ch}(E)$ & $I_{FB}=\int_M\hat{A}(S,F)\mathcal{CH}(S,F,E)$\\
\hline
$I_{BB}=\int_M\mathrm{ch}(\sigma(D))\frac{\mathrm{Td}(\mathrm{TM})}{e(\mathrm{TM})}$ & $I_{BB}=\int_M\mathcal{CH}(S,F,\sigma(D))\frac{\mathrm{Td}(S,F)}{\prod_i\lambda_i}$ \\
\hline
$I_{FF}=\int_M\hat{B}(\mathrm{TM})\mathrm{ch}(E)$& $I_{FF}=\int_M\hat{B}(S,F)\mathcal{CH}(S,F,E)$ \\
\hline
$I_{BF}=\int_M\mathrm{ch}(\sigma(D))\frac{\mathrm{Td}^*(\mathrm{TM})}{e(\mathrm{TM})}$ & $I_{BF}=\int_M\mathcal{CH}(S,F,\sigma(D))\frac{\mathrm{Td}^*(S,F)}{\prod_i\lambda_i}$ \\
\hline
\end{tabular}
\end{table}
\section{Conclusions and discussions}
In \cite{AG1983}, it was shown that considering quantum mechanical supersymmetric systems, specifically field theories in $0+1$ dimensions, and using $\text{Tr}(-1)^F$ as a topological invariant, one can compute the index density for the Atiyah-Singer index theorem for classical complexes \cite{AS1968a, AS1968b, AS1968c, AS1971a, AS1971b, ABP1973, Gilkey1974}, along with their corresponding G-index generalizations. At present, the index theorem derived from supersymmetry has become a central research framework in both theoretical and mathematical physics.

However, one might still ask: can non-supersymmetric (non-SUSY) physics also generate the Atiyah-Singer index theorem? In \cite{Li:2025qry}, we demonstrated that a simple example -- the quantum harmonic oscillator -- can lead to the generalized Hirzebruch signature theorem, a special case of the Atiyah-Singer index theorem. In this paper, we extend the results of \cite{Li:2025qry}. Using non-SUSY physics, specifically the Bose-Einstein and Fermi-Dirac distributions, we present a straightforward “physical” derivation of the Atiyah-Singer index theorem for vector bundles (finite dimensions) and sheaf (infinite dimensions). The case of infinite dimensions is far beyond modern mathmatics. In the non-degenerate case, where $e^{\alpha} \gg 1$, the four pairings “Fermi-Bose”, “Fermi-Fermi”, “Bose-Bose” and “Bose-Fermi” all reduce to the Euler class. In other words, the classical non-interacting system can give the Euler class. This represents a significant advance in mathematical physics, as it moves beyond the constraints of supersymmetry.

Beyond the specific derivations presented here, this work provides a profound and fundamental connection between quantum statistics and differential topology. We have demonstrated that this connection is not merely an analogy or a computational coincidence, but is rooted in a deep algebraic identity. The core structures of non-interacting quantum systems -- the Bose-Einstein and Fermi-Dirac distributions -- are intrinsically isomorphic to the generating functions of characteristic classes such as the Todd genus and the $\hat{A}$-genus. This isomorphism, formalized through the correspondence between the grand partition function $\Xi$ and the Chern character $ch$, reveals that the celebrated Atiyah-Singer index theorem can be derived from the principles of quantum statistics alone, without resorting to supersymmetry. Consequently, our findings suggest that certain topological invariants of manifolds and bundles are not solely geometric in origin; rather, in a very precise sense, they represent the universal quantum-statistical nature of matter. This establishes quantum statistics as a potentially more fundamental language for understanding these topological phenomena, offering a fresh and unifying perspective that bridges statistical mechanics, geometry, and topology at a much more fundamental level.

Indeed, readers may wonder whether there exists a unified framework capable of generating the Atiyah-Singer (AS) index theorem from both non-SUSY and SUSY physics. We believe that the answer is yes. We will explore this topic in a future paper.  

There are some promising research directions to explore. For example, what are the mathematical properties and physical meanings of the $\hat{B}$-genus for finite dimensions and infinite dimensions? Can we apply the pictures of statistical mechanics to understanding other mathematical theorems? Can the framework in Section 5 be connected to the established theorems? Can we apply this connection we discovered in this paper to studying physics, such as condensed matter physics? In short, this is an exciting research direction.








\end{document}